\documentclass[prl,twocolumn,reprint,showpacs,preprintnumbers,amsmath,amsthm,
amssymb,amsfont]{revtex4-1}
\pdfoutput=1
\usepackage{times}
\usepackage{graphicx}
\usepackage{dcolumn}
\usepackage{bm}
\usepackage{color}
\usepackage{subfigure}
\usepackage[version=4]{mhchem}
\providecommand{\angs}{\rm \AA}
\begin{document}

\title{Exploring the structural, electronic and magnetic properties of cation ordered 3$d$-5$d$
double perovskite Bi$_2$FeReO$_6$ and Bi$_2$FeIrO$_6$ thin-films from
first-principles}
\author
{Paresh C. Rout$^{(1)}$ and Varadharajan Srinivasan$^{(1,2)}$}
\affiliation{(1) Department of Physics, Indian Institute of Science Education
and Research Bhopal, Bhopal 462 066, India}
\affiliation{(2) Department of Chemistry, Indian Institute of Science Education
and Research Bhopal, Bhopal 462 066, India}
\begin{abstract}
We report a first-principles study of Bi-based 3$d$-5$d$ ordered double
perovskite oxides (\ce{A2BB$^\prime$O6}) with a 3$d$ atom (Fe) at the B-site and
5$d$ atoms (Re,Ir) at the B$^\prime$-site while keeping highly polarizable ions
(Bi$^{3+}$) at the A-site. We find that, under coherent heteroepitaxy, 
\ce{Bi2FeReO6} exhibits a strain-driven anti-ferromagnetic insulator to ferrimagnetic semi-metal transition, 
while \ce{Bi2FeIrO6} shows correlation driven ferromagnetic insulator to ferrimagnetic
half-metal transition with calculated magnetic moments of 5 $\mu_B$/f.u.
and 3 $\mu_B$/f.u., respectively. These properties along with the low band gaps in the insulating phases make the compounds appealing for spintronics applications. 
Furthermore, in \ce{Bi2FeIrO6}, the conduction and valence states are localized on different  transition metal 
sublattices implying more efficient electron-hole separation upon photoexcitation, a desirable feature for photovoltaic applications.
\end{abstract}

\pacs{
77.55.Nv, 75.50.Gg, 75.50.-i, 75.10.-b, 75.25.-j, 75.30.-m, 75.80.+q
}

\maketitle

\section{Introduction}
Transition metal (TM) double-perovskites (DP) (A$_{2}$BB$^\prime$O$_{6}$) have gained enormous
attention recently due to their fascinating properties in the areas of magnetism, ferroelectrics, spintronics
and multiferroics~\cite{vopson2015} and long served as a platform for device applications. Particularly, multiferroic materials
are quite promising due to their multipurpose technological
implications~\cite{spaldin2005,catalan2009}. These materials are characterized
by coexistence of ferroic orders such as ferroelectricity, ferromagnetism and/or
ferroelasticity along with a coupling of at least two of these orders which
can, in turn, lead to magnetoelectric, magnetoelastic effects, etc. A prominent
example of DP multiferroic is \ce{Bi2FeCrO6} (BFCO), which was designed from the
AFM parent multiferroic compound \ce{BiFeO3} (BFO) by replacing half the Fe ions by Cr 
with an aim to increase the magnetic moment through ferrimagnetic interaction of $\it{B}$ and $\it{B^\prime}$ sites
while retaining ferroelectricity and magneto-electric coupling
intact~\cite{baettig2005}. Due to AFM ordering, the parent compound BFO shows
$\sim$ 0 magnetic moment, however, BFCO was predicted to show a magnetic moment
of 2 $\mu_B$ per formula unit (f.u.) in its bulk form with a G-type AFM
ordering.

In practice, multiferroics can be prepared as high quality epitaxial
thin films on oxide substrates. The epitaxial strain (ES), which can be
introduced by varying lattice mismatch between the thin-film and substrate, has
the tendency to control the material properties and to stabilize other
metastable magnetic structures~\cite{meng2018strain,james,strain2,Millis_PRB_2016,gan1998}.
The potential of multiferroic materials towards photovoltaic (PV) and photocatalytic (PC) application
is currently being explored. The narrow band gap, which arises due to electron-electron interaction
governing the magnetic ordering, makes multiferroics special candidates for PV application over
the general perovskite materials~\cite{grinberg2013}. Among multiferroics, BFO 
was considered to show appreciable PV effect due to the presence
of a direct band gap, a fact latter corroborated by Chui \textit{et al.}~\cite{choi2009}. However,
the power conversion efficiency (PCE) of BFO based thin film devices is still very low for
practical application. Lowering the band gap of these oxides
without affecting their FE properties is a promising conceptual
route to obtain solar energy conversion devices with higher PCE. Recently, it
was shown that the epitaxial multiferroic ordered BFCO thin-films possess a
power conversion efficiency of 6.5\% under the illumination of red
light laser~\cite{nechache2011} and 8.1\% under AM1.5G
illumination~\cite{nechache2015}. In both the works, it was pointed out that the
Fe and Cr cation order and the narrow band gap play a crucial role in the
PV performance of BFCO thin films. Kim \textit{et al.}~\cite{kim2018} recently attributed 
the excellent performance of BFCO to the spatial separation of the photoexcited electron
and hole states onto the Fe and Cr sites, respectively, with the extent of separation increasing 
with the disorder at the $B$-site. 

Recently, we have shown that epitaxially grown (001) BFCO thin-films are
unstable to anti-site defects and prefer a C-type AFM (C-AFM) ordered ground
state~\cite{Paresh}, both features leading to loss of magnetization~\cite{khare1,shabadi,khare2}.
The anti-site defects (disordering) occur basically due to the similarity of charge and ionic radii of Fe$^{3+}$
and Cr$^{3+}$ ions. We have also recently shown that by manipulating both ES and
aliovalent A-site doping in BFCO thin films~\cite{rout2018epitaxial}, one can not only mitigate the issues
like cation disorder and low magnetism but also significantly reduce the band gap, a desirable feature for PV
applications. An alternative approach to suppress B-site disorder and encode functionality in Bi-based DPs is 
to form 3$d$-5$d$ oxides~\cite{MarjanaPRB}. Doping with 5$d$ TM ions can not only improve magnetisation but also prevent the
formation of antisite defects like BFCO thin-films due to their larger sizes compared to the 3$d$ ions. Apart from this, the mixture of
3$d$-5$d$ ions would likely improve the transition temperature as seen in case of Ca, Sr-based systems, due to the induced magnetic moment at the 
nonmagnetic sites~\cite{feng2014high,HighTc}.

In this work, we investigate the strain-dependent magnetic and electronic properties of two new double-perovskite 3$d$-5$d$ systems: \ce{Bi2FeReO6} (BFRO) and \ce{Bi2FeIrO6} (BFIO) 
from first-principles DFT simulations. BFO was chosen as a starting point for the doping with 5$d$ ions given its robust ferroelectric nature. We chose
Fe atom as the 3$d$ ion in the B-site as it usually occurs in high-spin 3+ oxidation state, which could lead to large magnetic moment in DPs, stronger magnetic interactions and hence larger transition temperatures.

Our calculations predict an anti-polar $P2_1/n$-like structure for both compounds 
if grown on cubic substrates. We find that a C-type AFM ordering is favored for BFRO, while
BFIO becomes a ferromagnetic insulator through out the ES region. 
%%%%%%%%%%%%%%%%%%%%%%%%%%%%%%%%%%%%%%%%%%%%%%%%%%%%%%%%%%%%%%%%%%%%%%%%%%%%%
\begin{figure}[pbht!]
\includegraphics[trim=0.0cm 1.0cm 0.0cm
0.0cm,clip=true,width=0.48\textwidth]{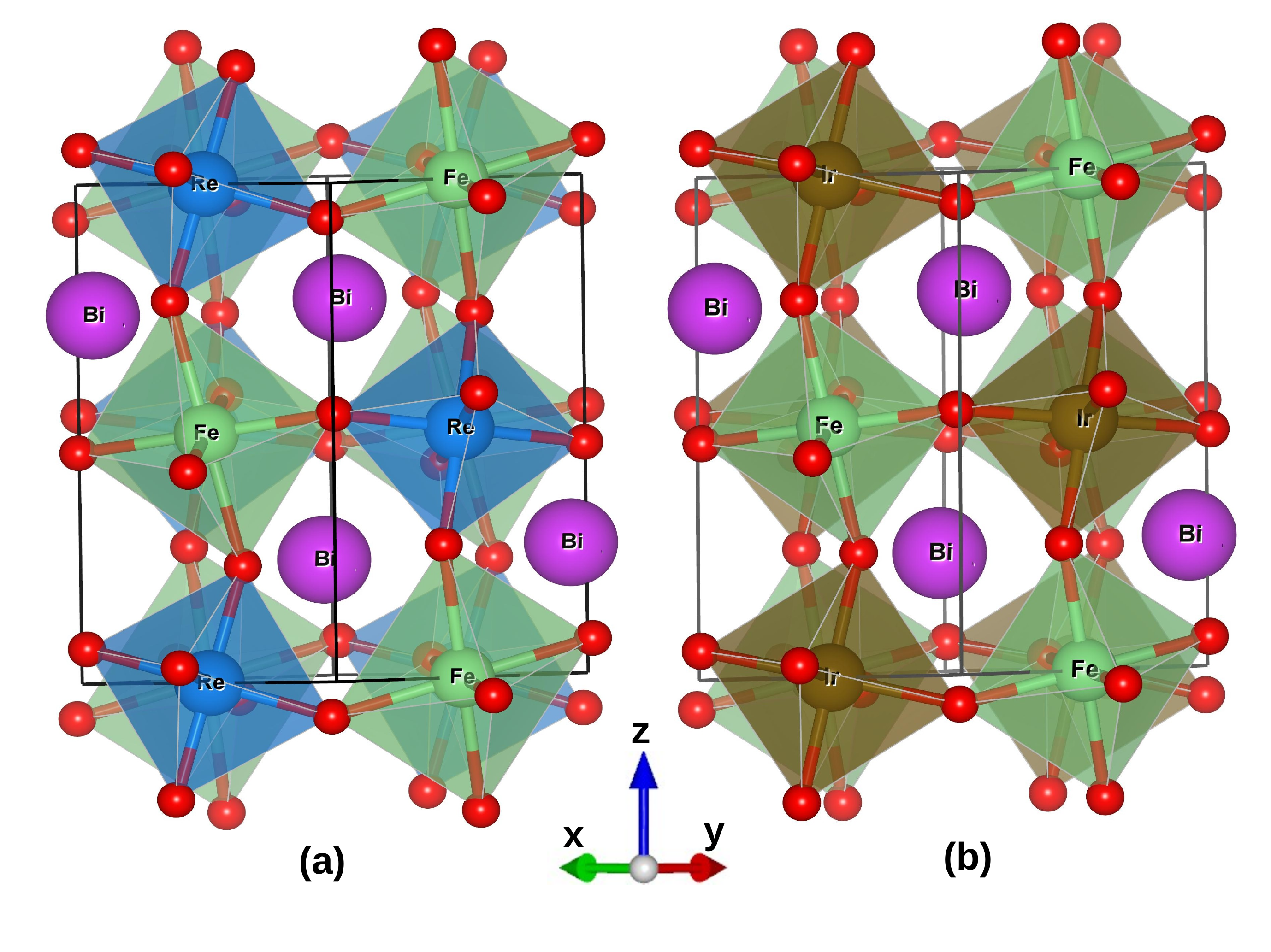}
\caption{Strained bulk tetragonal supercell used for modeling the thin-films
(001) for calculations. (a) The tetragonal supercell of
\ce{Bi2FeReO6} with $a^-a^-c^+$ oxygen octahedra tilt pattern, (b) Represents,
the tetragonal supercell of \ce{Bi2FeIrO6} with same octahedra orientation as
indicated in (a). Both the structure has $P2_1/n$ symmetry.}
\label{Structure_3d-5d}
\end{figure}
%%%%%%%%%%%%%%%%%%%%%%%%%%%%%%%%%%%%%%%%%%%%%%%%%%%%%%%%%%%%%%%%%%%%%%%%%%%%%%
In BFRO thin-films, we show that at higher compressive strain, there
is a C to G-AFM (1$\mu_B$/f.u.) transition accompanied by monoclinic to monoclinic
structural transition due to octahedral reorientation. However, in
BFIO thin-films, we find a correlation driven metal-to-insulator
transition under ES. We find that the 3$d$ and 5$d$ cations adopt high spin 2+ and 4+ oxidation
state (denoted below as (2+,4+)), respectively, in case of BFRO thin films. In BFIO, however,
there is a competition between (2+,4+) and (3+,3+) states affected by onsite correlation at the low-spin Ir sites.
We predict that correlation effects along with the stability of the 3+ oxidation state on the high-spin Fe sites 
will lead to the stabilisation of the (3+,3+) state in BFIO, with a magnetic moment of 5$\mu_B$/f.u., under all ES.
Furthermore, BFRO is predicted to be an AFM insulator under ES, whereas the
Ir-based DP is a ferromagnetic insulator. Ferromagnetic insulators (FM-I) are
rare and very promising for their spintronics and PV applicability~\cite{meng2018strain,LVO}.
So far, most of the synthesized 3$d$-5$d$ DP materials
are half-metallic in nature. However, \ce{Bi2FeIrO6} is a promising
material as it exhibits both ferromagnetism and insulating properties
simultaneously. Considering the presence of 5$d$ elements, we included SOC in our 
calculation but found that SOC does not have a significant effect on the magnetic properties while it has a
small impact on the electronic properties, particularly resulting in a slight band gap reduction.
We show that the correlation effect on 5$d$ site plays a very crucial in opening the band gap and magnetic phase transition,
which was completely ignored in the previous study~\cite{Marjana}. Presence of
very narrow band gap $\sim$1.1 eV in both the compounds (considering correlation
effect on both 3$d$-5$d$ ions) makes them suitable candidates for PV
applications. 

\section{Method of Calculation}
Our calculations employed a spin-polarized
GGA+$U$~\cite{anisimov,matteo,wentzco} approach using
the revised version Perdew-Burke-Ernzerhof (PBE)~\cite{pbe1,pbe2} for solids, PBEsol~\cite{PBEsol} as
exchange-correlation functional within the framework of the Quantum-ESPRESSO
code~\cite{paulo}. Ionic cores were modeled by ultrasoft pseudopotentials keeping the 4$f$ electrons of 5$d$ transition metals
and Bi atoms as part of the core. Plane-wave cut-off energy of 85 Ry was used for representing wavefunctions, and 700 Ry 
for the augmentation charge.  The Hubbard $U$ parameter can be calculated {\it ab initio} using the linear response formalism
~\cite{matteo,wentzco}. We found the $U$ values to be 6.4 eV (Fe) and 3.6
eV (Re) in BFRO system, whereas in BFIO case, the $U$ values were 6.38 (Fe) and 5.18 (Ir). 
We used a $10\times10\times 8$ Monkhorst-Pack k-point mesh for Brillouin zone integration. For the density of
states (DOS) calculations, a denser Monkhorst-Pack k-point grid of
$16\times 16\times 16$ was employed. All the structures were relaxed until the
Hellmann-Feynmann forces are less than 0.26 meV/{\AA}. 

For both the compounds, we have constructed 20-atom
$\sqrt{2}\times\sqrt{2}\times2$ tetragonal supercells (shown in
Figure~\ref{Structure_3d-5d}), starting from a simple cubic double-perovskite
structure, to allow for appropriate magnetic ordering of ions along (111)
direction. The structures were chosen to conform to $P2_1/n$ (antipolar) and
$R$3 (polar) space groups, respectively, given their compatibility with cubic substrates~\cite{khare1,khare2}.
These essentially differ by the sense and relative magnitudes of oxygen octahedral rotations as indicated by the
Glazer notations $a^- a^-c^+$ and $a^- a^- a^-$, respectively~\cite{glazer1975,glazer1972}.  Using the substrate
\ce{SrTiO3} pseudo-cubic lattice parameter $a_{cub}=3.90$\angs~~\cite{sto} as a reference, we generated
structures mimicking the epitaxially strained films by varying the in-plane
lattice parameters over a range of realistic substrate strains corresponding to
(001) epitaxial growth. The in-plane lattice parameters of supercell are
set to $\bar{a}=\bar{b}=\sqrt{2}\times a_{cub}$ while the $c$ parameters have
been relaxed for each in-plane strain. The ES can be defined as 
\begin{eqnarray}
\boxed{\epsilon=(\bar{a}
-\bar{a}_{ref})/\bar{a}_{ref}}
\label{strain_Re-Ir}
\end{eqnarray} where $ \bar{a}_{ref} $ is the unstrained
lattice parameter. The selection of Re and Ir structures is based on
the adapted tolerance factor $\bf t$ of the
double-perovskites~\cite{tollerance}. The tolerance factor for DPs
A$_{2}$BB$^\prime$O$_6$, can be written as:  
\begin{eqnarray}
\boxed{\bf{t=\frac{r_A+r_O}{\sqrt2\times(\frac{r_{B}}{2}+\frac{r_{B^\prime}}{2
} +r_O)}}}
\label{toll3d-5d}
\end{eqnarray}
where r$_A$, r$_B$ and r$_B^\prime$ are the ionic radii of the respective ions
and r$_O$ is the ionic radius of oxygen. It is known that, for
double-perovskite family, if t $<$ 0.97 the compound becomes either monoclinic
($P2_1/n$) or orthorhombic~\cite{Serrate1}. In case of \ce{Bi2FeReO6}, the
calculated average tolerance factor is 0.93 while for \ce{Bi2FeIrO6}, it is
0.95. Hence, both structures are expected to take perovskite form. 

\section{Results and Discussions}
\subsection{Strain-dependence ground-state structure of \ce{Bi2FeReO6}}
We first investigated the effect of ES on the ground-state structure
of BFRO compound. We considered three types of AFM (A, C, G-type) and the FM
orders for the calculations~\cite{Paresh}. We have considered the two lowest
energy phases $P2_1/n$ and $R$3 phase, which were also previously found to be
stable for similar systems under various strains ~\cite{Marjana}. The ES
is introduced in thin-film by constraining the two in-plane lattice vectors to
be equal in length and enforce the angle between them to be 90$^\circ$.
%%%%%%%%%%%%%%%%%%%%%%%%%%%%%%%%%%%%%%%%%%%%%%%%%%%%%%%%%%%%%%%%%%%%%%%%%%%%%%%
%\vspace{-0.8cm}
\begin{figure}[pbht!]
\includegraphics[trim=0.0cm 1.0cm 0.0cm
2.5cm,clip=true,width=0.50\textwidth]{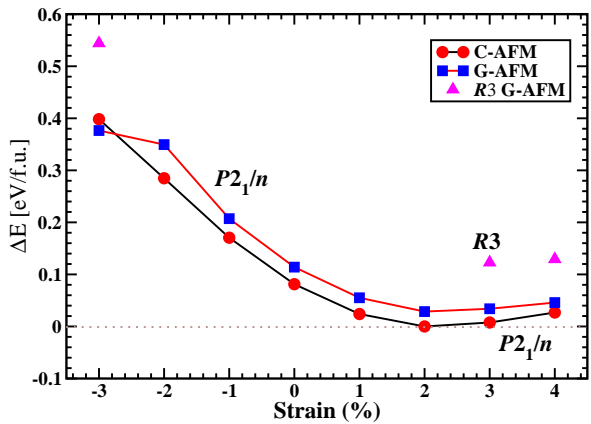}
\caption{Energy vs strain for the two lowest energy magnetic ordering
of $P2_1/n$ structure. The energy of $R$3 structure indicated by magenta upper
triangle are plotted at at higher compressive and tensile strain region.
All the energies are positioned with respect to the lowest energy of C-AFM
ordering in $P2_1/n$ symmetry. A crossover between G-AFM and
C-AFM ordering occurs at -2.7\% strain.}
\label{StrainvsEnergy2}
\end{figure}
%%%%%%%%%%%%%%%%%%%%%%%%%%%%%%%%%%%%%%%%%%%%%%%%%%%%%%%%%%%%%%%%%%%%%%%%%%%%%%%%
%%%%%%%%%%%%%%%%%%%%%%%%%%%%%%%%%%%%%%%%%%%%%%%%%%%%%%%%%%%%%%%%%%%%%%%%%%%%%%%%
We first obtained the optimized geometries in the $P2_1/n$ and $R$3 phases for
different magnetic orderings at all the considered ES. The minimum energy
structure at each strain was obtained by allowing the length and
angle of out-of-plane lattice parameter to relax simultaneously along with the
ionic positions. Figure~\ref{StrainvsEnergy2} depicts the energy (relative to
the global minimum structure, i.e. $P2_1/n$ at 2\% strain) of the lowest lying
states as a function of ES. From Figure~\ref{StrainvsEnergy2}, it
is clear that the ferroelectric phase ($R$3) is not favorable under the
considered epitaxial strains. 
%%%%%%%%%%%%%%%%%%%%%%%%%%%%%%%%%%%%%%%%%%%%%%%%%%%%%%%%%%%%%%%%%%%%%%%%%%%%%%%
%%%%%%%%%%%%%%%%%%%%%%%%%%%%%%%%%%%%%%%%%%%%%%%%%%%%%%%%%%%%%%%%%%%%%%%%%%%%%%%
%\vspace{-0.9cm}
\begin{figure}[pbht!]
\includegraphics[trim=0.0cm 0.5cm 0.0cm
2.0cm,clip=true,width=0.450\textwidth]{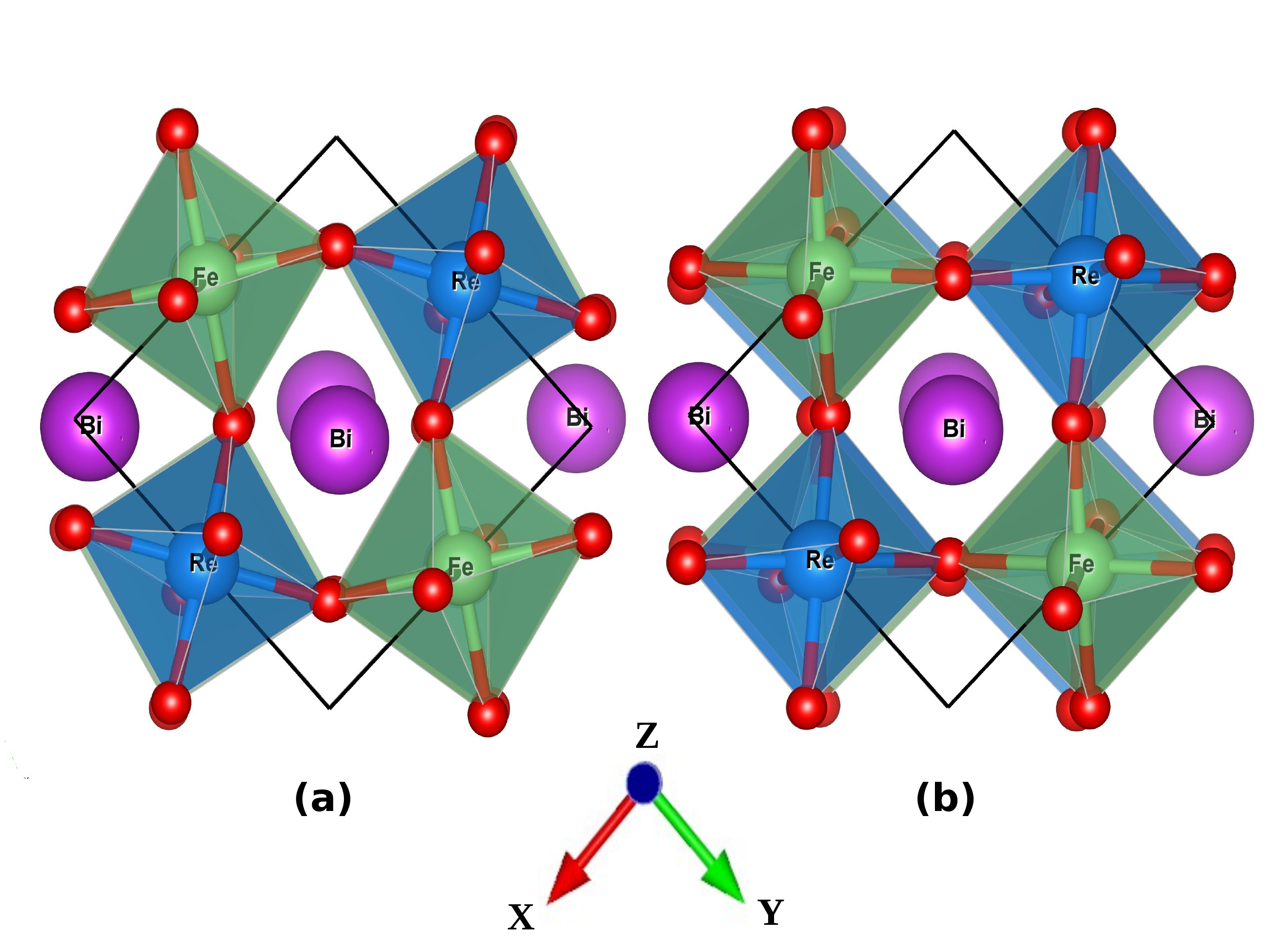}
\caption{(a) Out-of-plane (top) view of G-AFM \ce{Bi2FeReO6} structure,
(b) Out-of-plane (top) view of C-AFM \ce{Bi2FeReO6} structures
at -2.7\% strain.}
\label{C-type_G-type}
\end{figure}
%%%%%%%%%%%%%%%%%%%%%%%%%%%%%%%%%%%%%%%%%%%%%%%%%%%%%%%%%%%%%%%%%%%%%%%%%%%%%%%
We find that C-AFM of $P2_1/n$ phase is the most stable structure over the
broad range of strain leading to zero total magnetic moment due to exact
cancellation of the spin magnetic moment. 
%This result has been confirmed by a
%very recent experiment~\cite{Rana}. 
Most interestingly, a magnetic ordering
transition (from C-AFM to G-AFM) is seen at -2.7\%, which suggests
that a net magnetic moment can also be achieved in \ce{Bi2FeReO6} thin-films
through compressive strain. We ascribe the magnetic phase transition to
the oxygen octahedra reorientation induced monoclinic to monoclinic structural
transition. The structures of both C and G-AFM BFRO are shown in
Figure~\ref{C-type_G-type} and the corresponding structural information for
both the magnetic ordered structure at -2.7\% strain are provided in
Table~\ref{tableC-G}. 
%%%%%%%%%%%%%%%%%%%%%%%%%%%%%%%%%%%%%%%%%%%%%%%%%%%%%%%%%%%%%%%%%%%%%%%%%%%%%%%%
\begin{table}[h!] 
\caption{Structural information for both C-AFM and G-AFM magnetic ordering at 
-2.7\% strain.}
\label{tableC-G}
\begin{center}
\begin{tabular}{l|l|l|l|l|l|l}
\hline
\hline 
 Type & $a$ (\AA) & $b$ (\AA) & $c$ (\AA) &  $\alpha (^\circ) $& $\beta (^\circ) $ &
$\gamma (^\circ)$ \\
\hline
 C-AFM  & 5.364 & 5.364 & 8.652 &  90.00 &  90.28 &  90.00 \\
\hline
 G-AFM  & 5.364 & 5.364 & 9.094 &  90.00 &  84.75 &  90.00 \\
\hline \hline 
\end{tabular}
\end{center}
\end{table}
%%%%%%%%%%%%%%%%%%%%%%%%%%%%%%%%%%%%%%%%%%%%%%%%%%%%%%%%%%%%%%%%%%%%%%%%%%%%%
The monoclinic angle (angle between $a$ and $c$ lattice parameters) changes from
being obtuse in the C-AFM to be acute in the G-AFM structure. This
transformation results from a sudden oxygen octahedra reorientation at
-2.7\% strain (see Figure~\ref{C-type_G-type}). So, beyond -2.7\% strain, 
magnetic ground-state for BFRO can be achieved with a magnetic moment of
1$\mu_B$/f.u.. The polar $R$3 phase, however, remains higher in energy in this
strain region. One unusual finding in case of BFRO thin-film is the oxygen octahedral
distortion around the Fe atom. It was observed that the Fe atom undergoes a Jahn-Teller-like distortion, leading
to three different types of Fe-O bond length whereas the Re atom remains
undisturbed. The three unequal bonds at reference (0\% strain) are axial
Fe-O$_{ax}$ long bond (2.28 \AA), equatorial Fe-O$_{eq}$ medium bond (2.14 \AA)
and apical Fe-O$_{ap}$ short bond (2.05 \AA). This octahedral distortion
pattern is also seen in other strain regions. 
%%%%%%%%%%%%%%%%%%%%%%%%%%%%%%%%%%%%%%%%%%%%%%%%%%%%%%%%%%%%%%%%%%%%%%%%%%%%%%%%
%%%%%%%%%%%%%%%%%%%%%%%%%%%%%%%%%%%%%%%%%%%%%%%%%%%%%%%%%%%%%%%%%%%%%%%%%%%%%%%%
\subsection{Electronic and magnetic properties of \ce{Bi2FeReO6}}
In this section, we analyze the electronic structure of \ce{Bi2FeReO6}
thin-films in the antipolar $P2_1/n$ structure. We also briefly discuss about
the magnetic properties of the corresponding structure.

\begin{figure*}[pbht!]
\centering
\subfigure[]
{
\includegraphics[trim=0.0cm 0.0cm 0.0cm
2.0cm,clip=true,width=0.48\textwidth]{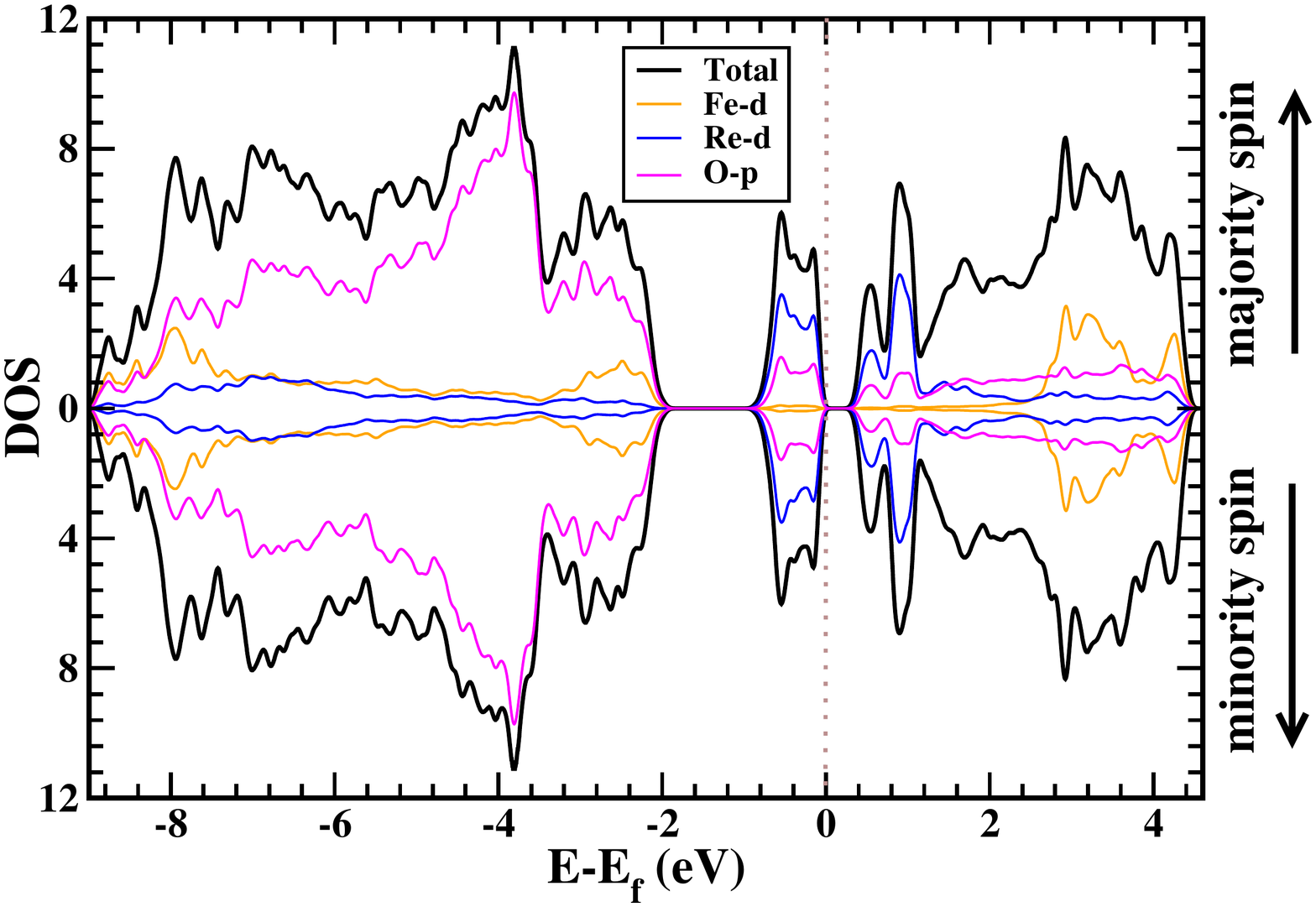}
\label{BiFeRe_DOS1}
}
\subfigure[]
{
\includegraphics[trim=0.0cm 0.0cm 0.0cm
2.0cm,clip=true,width=0.48\textwidth]{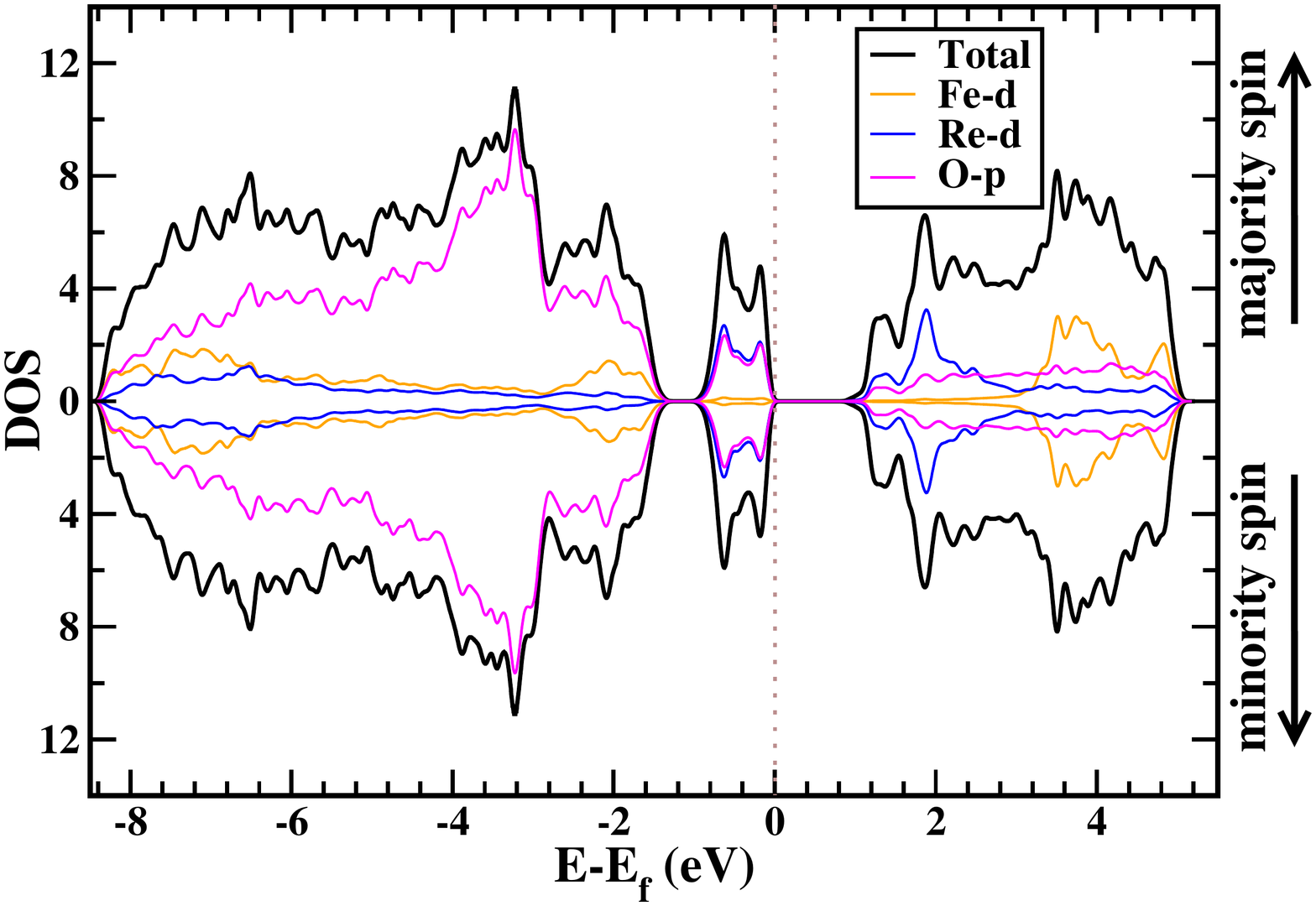}
\label{PDOS_C-type_both}
}
\caption{(a) Density of states (DOS) plot of ground-state \ce{Bi2FeReO6}
structure by adding $U$ only on Fe-site. The solid black line,
total DOS for both majority and minority spin channels; orange line, Fe-$d$
orbitals; blue solid lines Re-$d$ orbitals and solid magenta for O-$p$ states.
The local magnetic moments on Fe and Re are exactly antialigned. (b) Depicts DOS
for \ce{Bi2FeReO6} by adding on-site Coulomb repulsion $U$ on both Fe and
Re-site. The color code has the same meaning as mentioned in (a). The band gap
enhanced up to 1.2 eV, making the system into a robust insulator.}
\label{}
\end{figure*}

We find that the BFRO structure prefers to adopt C-AFM ordering, where
the spins on Fe (3$d$) and Re (5$d$) transition metal ions anti-aligned in the plane of 
epitaxy (in-plane) but align parallelly perpendicular to this plane (out-of-plane). 
However, the structure undergoes a magnetic phase transition beyond
-2.7\% strain leading to a G-AFM ordered phase where the spins are anti-aligned both in-plane and out-of-plane. 
The total density of states (DOS) and the contributions to it from the $d$ and $p$ states of Fe (Re) and O, respectively, for
both the spin channels, calculated within GGA+$U$ (only on Fe) at reference
strain (0\%) are shown in Figure~\ref{BiFeRe_DOS1}.
%%%%%%%%%%%%%%%%%%%%%%%%%%%%%%%%%%%%%%%%%%%%%%%%%%%%%%%%%%%%%%%%%%%%%%%%%%%%%%%%
%%%%%%%%%%%%%%%%%%%%%%%%%%%%%%%%%%%%%%%%%%%%%%%%%%%%%%%%%%%%%%%%%%%%%%%%%%%%%%%%
In BFRO, Re prefers 4+ oxidation state (high spin) and Fe prefers to stay in high spin 2+ oxidation
state. The thick black line represents the total DOS for both the spin channels
(spin up and down). The up and down channels contribute equally to the DOS indicating an AFM ground state. The
orange solid line represents the local density of states (LDOS) for Fe (3$d$)
orbits and the solid blue lines LDOS for 5$d$ (Re) states. The LDOS which arise
from oxygen $p$ orbitals, denoted by magenta color, contribute predominantly 
to the total density of states (see Figure~\ref{BiFeRe_DOS1}). The structure is
insulating, with a small gap of 0.4 eV in the GGA+$U$ calculations. However, this gap is significantly enhanced up to
1.2 eV after incorporating onsite Coulomb repulsion $U=$3.6 eV on Re-sites as shown in
Figure~\ref{PDOS_C-type_both}. 
%%%%%%%%%%%%%%%%%%%%%%%%%%%%%%%%%%%%%%%%%%%%%%%%%%%%%%%%%%%%%%%%%%%%%%%%%%%%%%%%
This suggests that the correlation effects arising from Re atoms would be important in this system.
The band near the Fermi energy arises predominantly from
Re 5$d$ ($t_{2g}$) states with strong oxygen 2$p$ orbital hybridization. The
bands just above the Fermi level have a predominantly Re 5$d$ ($e_g$) character
in both spin channels. The Fe $t_{2g}$ bands lie between -8.5 to -5 eV for both the spin
channels while the $e_g$ bands extend from -5 to about -1.8 eV below the Fermi
energy. There are two in-plane and one out-of-plane Fe-O-Re
angles measuring 141$^\circ$, 149$^\circ$ and
148$^\circ$, respectively. The two in-plane angles are different due to oxygen octahedral
distortion around the Fe atom. 
%%%%%%%%%%%%%%%%%%%%%%%%%%%%%%%%%%%%%%%%%%%%%%%%%%%%%%%%%%%%%%%%%%%%%%%%%%%%%%%%
\begin{figure}[pbht!]
\includegraphics[trim=0.0cm 1.0cm 0.0cm
2.5cm,clip=true,width=0.48\textwidth]{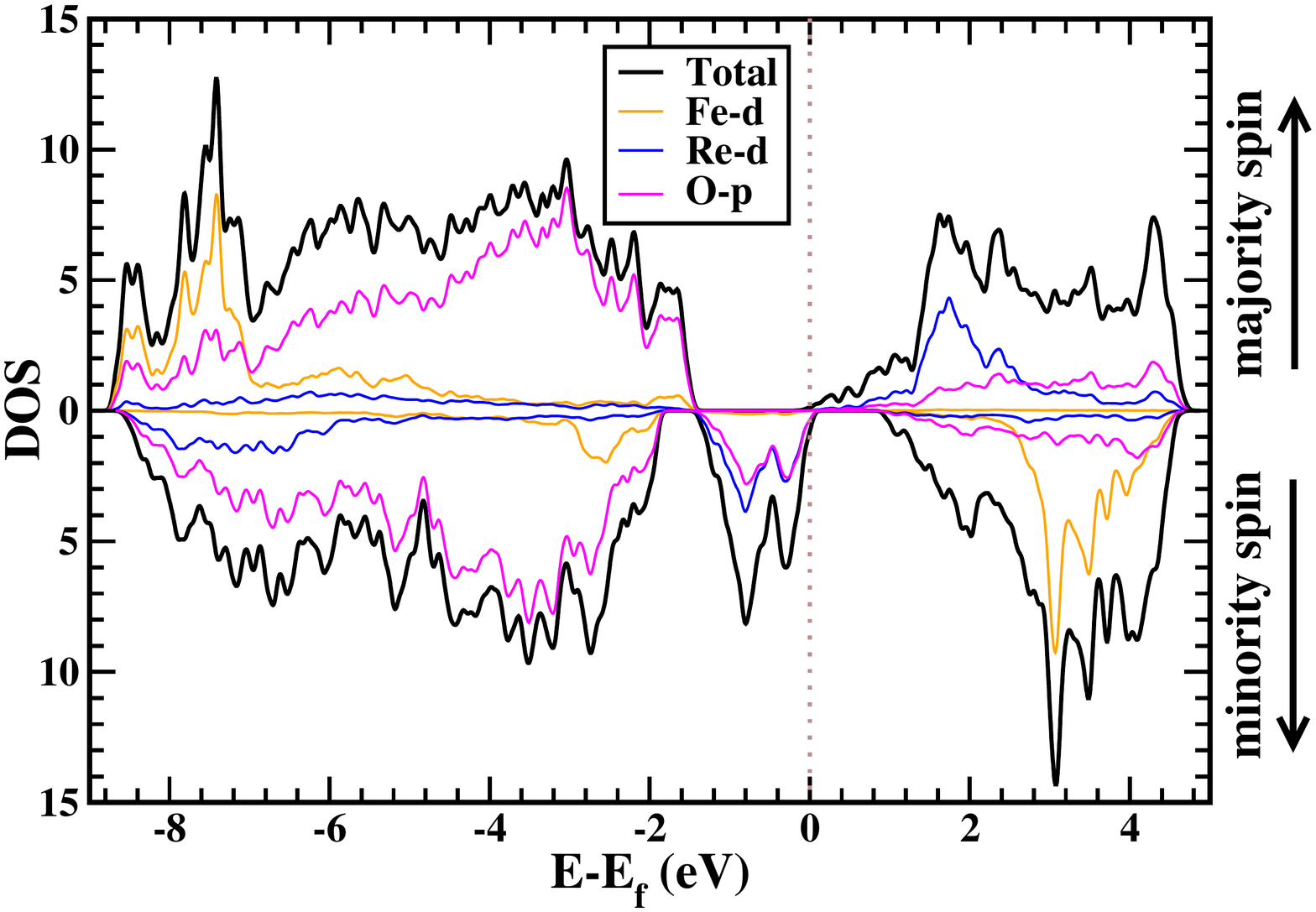}
\caption{The density of states (DOS) of G-type AFM \ce{Bi2FeReO6} at
-2.7\% strain calculated using GGA+$U$ method with $U_{eff}$=6.4, 3.6 eV for
Fe and Re, respectively. The dashed vertical zero line is set to the Fermi
energy. The local magnetic moments on Fe and Re are antialigned with unequal
magnitude leading to a net magnetic moment on the system. The solid black
line, total DOS; orange solid line, Fe-$d$ states; blue line, Re-$d$ states and
solid magenta line, O-$p$ states.}
\label{PDOS_G-type}
\end{figure}
%%%%%%%%%%%%%%%%%%%%%%%%%%%%%%%%%%%%%%%%%%%%%%%%%%%%%%%%%%%%%%%%%%%%%%%%%%%%%%%%
With these angular orientations, the magnetic moments on Fe and Re are coupled
through antiferromagnetic superexchange mechanism~\cite{good,kanamori1,Wollan}
along the in-plane direction and coupled ferromagnetically along out-of-plane.

Similarly, the DOS for G-AFM at -2.7\% strain is also shown in
Figure~\ref{PDOS_G-type}. Here, the correlation effect on both Fe and
Re-sites are taken into account for the DOS calculations. Unlike the
C-AFM phase, in this case, neither the total DOS nor the LDOS are identical
in both the spin channels. The difference in the DOS indicates the ferrimagnetic (FiM)
nature of the compound. The $t_{2g}$ bands of Fe are sharply localized within
-8.5 to -6.0 eV energy window, while the $t_{2g}$ states of Re-sites are
localized in the minority spin channel immediately below the Fermi energy. 
Interestingly, the valence and conduction band edges correspond to different spin channels
and there is no gap in between. Hence, there is an insulator (C-AFM) to semi-metal (G-AFM) transition at -2.7\%
strain. The half-metallic (or semi-metallic) state (see Figure~\ref{PDOS_G-type}) results despite using $U$ on both the TM sites,
making it less likely to be an artifact of the DFT method employed. This is indeed one of our important findings in this compound under ES. 

\subsection{Strain-dependence ground-state structure of \ce{Bi2FeIrO6}}
\begin{figure*}[pbht!]
 \centering
\subfigure[]
{
\includegraphics[trim=0.0cm 1.0cm 0.0cm
2.0cm,clip=true,width=0.48\textwidth]{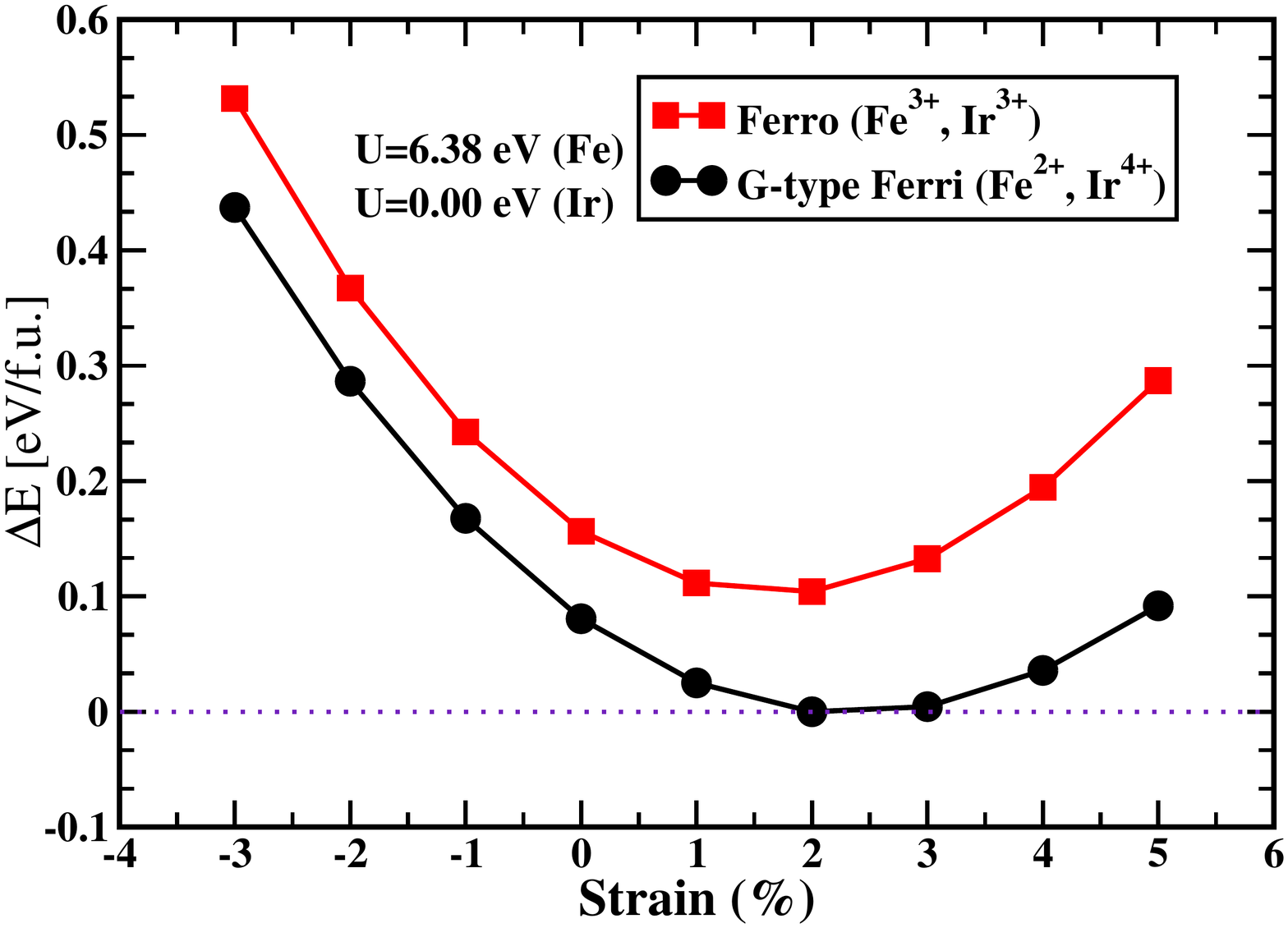}
\label{EVsStr_Fe}
}
\subfigure[]
{\includegraphics[trim=0.0cm 1.0cm 0.0cm
2.0cm,clip=true,width=0.48\textwidth]{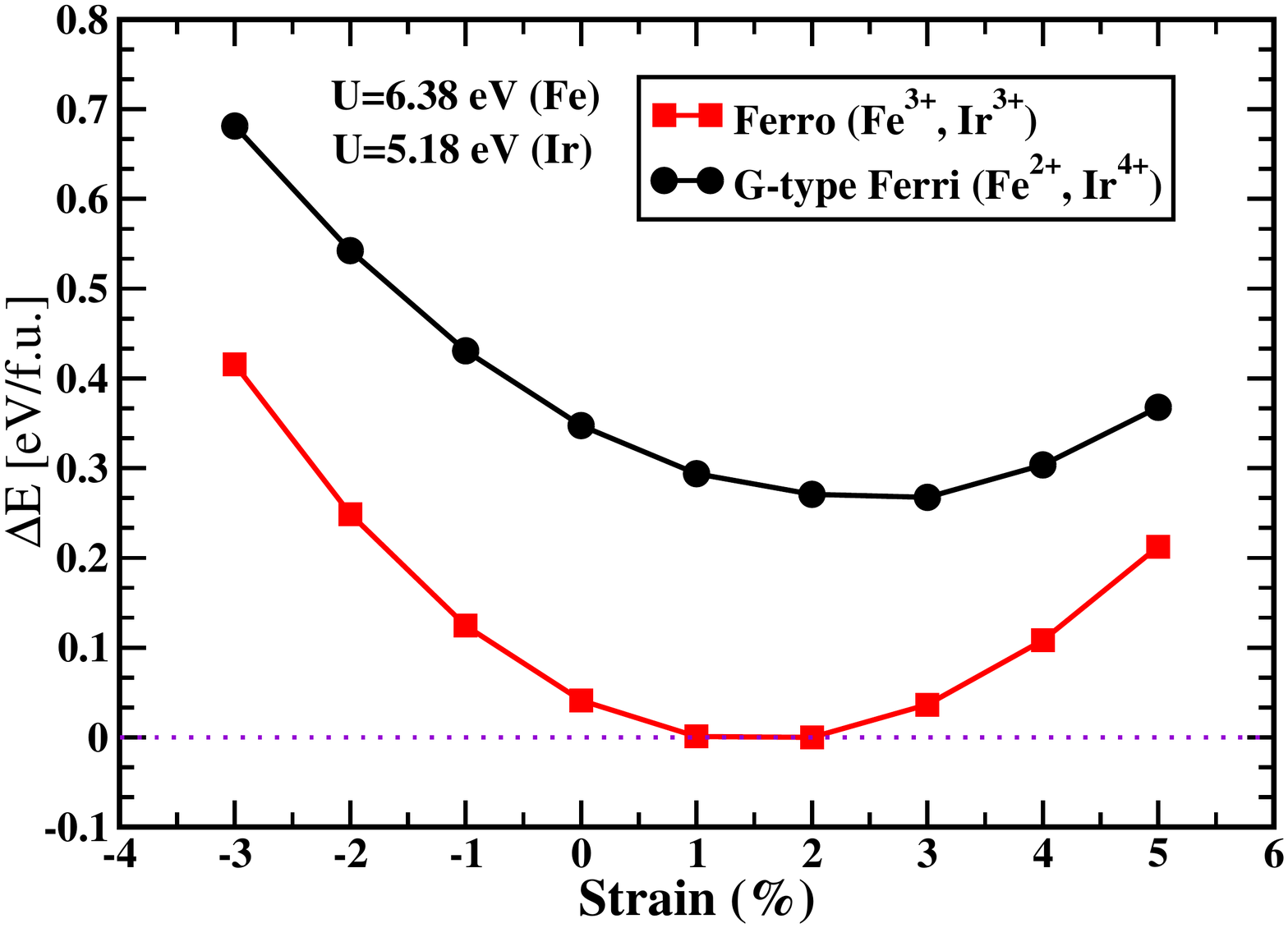}
}
\label{EVsStr_Fe-Ir}
\caption{~(a) Strain dependence structural stability of \ce{Bi2FeIrO6} thin
films. Optimized energies are positioned as a function of biaxial with respect
to the G-type ferrimagnetic structure at +2\% strain. All the energies were
optimized using GGA+$U$ (Fe) approach. The solid black line denotes the
ferrimagnetic metallic structure and the solid red line represents the
ferromagnetic insulator.~(b) Illustrates the energy \textit{vs} biaxial strain 
where the energies are positioned with respect to the ferromagnetic
structure at +2\% strain. The energies were obtained using the GGA+$U$ method
with $U_{eff}$= 6.38, 5.18 eV for Fe and Ir, respectively. Like
Figure~\ref{EVsStr_Fe}, the solid black line denotes the ferrimagnetic metallic
state and that of the solid red line denotes the ferromagnetic insulator, as the
lowest energy state.}
\label{EVsStr}
\end{figure*}

In this section, we have studied the influence of ES on
\ce{Bi2FeIrO6} thin-films. The ionic positions and the corresponding $c$-axis
at each strain are optimized by using GGA+$U$ (used $U$ only on
Fe) approach in $P2_1/n$ symmetry. We also studied the effect of $U$ on Ir (5$d$) sites as it plays a
very crucial role in predicting the electronic properties as seen in the previous section. The optimized energies
are plotted against the ES as shown in Figure~\ref{EVsStr} which shows the
lowest energy magnetic states with large magnetic moments due to the existence of different oxidation numbers in the TM ions.
As can be seen in Figure~\ref{EVsStr_Fe}, the G-AFM emerged as the lowest energy state
throughout the epitaxial strain. The FiM ground-state (solid black pointed circle
line) possesses a 2+ (Fe) and 4+ (Ir) oxidation state which contributes to a net
magnetic moment of 3 $\mu_B$/f.u., while the solid square pointed line represents the
higher energy ferromagnetic structure with 3+ (Fe) and 3+ (Ir) oxidation states.
In both the cases, the Fe-site assumes a high-spin configuration while Ir prefers low-spin. 
%%%%%%%%%%%%%%%%%%%%%%%%%%%%%%%%%%%%%%%%%%%%%%%%%%%%%%%%%%%%%%%%%%%%%%%%%%%%%
%%%%%%%%%%%%%%%%%%%%%%%%%%%%%%%%%%%%%%%%%%%%%%%%%%%%%%%%%%%%%%%%%%%%%%%%%%%%%
%%%%%%%%%%%%%%%%%%%%%%%%%%%%%%%%%%%%%%%%%%%%%%%%%%%%%%%%%%%%%%%%%%%%%%%%%%%%%%%%
%%%%%%%%%%%%%%%%%%%%%%%%%%%%%%%%%%%%%%%%%%%%%%%%%%%%%%%%%%%%%%%%%%%%%%%%%%%%%%%%
Note that the energy \textit{vs} strain (Figure~\ref{EVsStr_Fe}) was obtained by
using the GGA+$U$ approach, where the on-site Coulomb repulsion only included for
Fe-site. However, keeping the necessity of $U$ values on 5$d$-sites in
mind~\cite{Hua}, when we added $U$ (5.18 eV) on Ir sites, we observed a sudden
change in the ground state. Moreover, adding $U$ on Ir sites the FiM 
(${2+}$, ${4+}$) state no longer remains as the
lowest energy state. Instead, the ferromagnetic (${3+}$, ${3+}$) state get
stabilized under strain (see Figure~\ref{EVsStr_Fe-Ir}). The FiM
(${2+}$, ${4+}$) state denoted by solid black line is shifted up in
energy while the energy of the (${3+}$, ${3+}$) state, represented by the solid red line, is
brought down significantly as compared to Figure~\ref{EVsStr_Fe}.
%%%%%%%%%%%%%%%%%%%%%%%%%%%%%%%%%%%%%%%%%%%%%%%%%%%%%%%%%%%%%%%%%%%%%%%%%%%%%%%%
%%%%%%%%%%%%%%%%%%%%%%%%%%%%%%%%%%%%%%%%%%%%%%%%%%%%%%%%%%%%%%%%%%%%%%%%%%%%%%%%
The ferromagnetic ground-state leads to a total magnetic moment of 5
$\mu_B$/f.u., which comes predominantly from the Fe sites, while the Ir ion
undergoes a low spin $d^6$ configuration. In the Bi-based 3$d$-5$d$ compounds,
this is a first observation of a 3+ (3$d$), 3+ (5$d$) states. In this compound,
the magnetic moments of Fe atoms are aligned ferromagnetically and the induced
moments on Ir sites are almost zero due to the low spin configuration as a
result of which the total system becomes ferromagnetic. 

\begin{figure}[pbht!]
\includegraphics[trim=0.0cm 1.0cm 0.0cm
2.0cm,clip=true,width=0.48\textwidth]{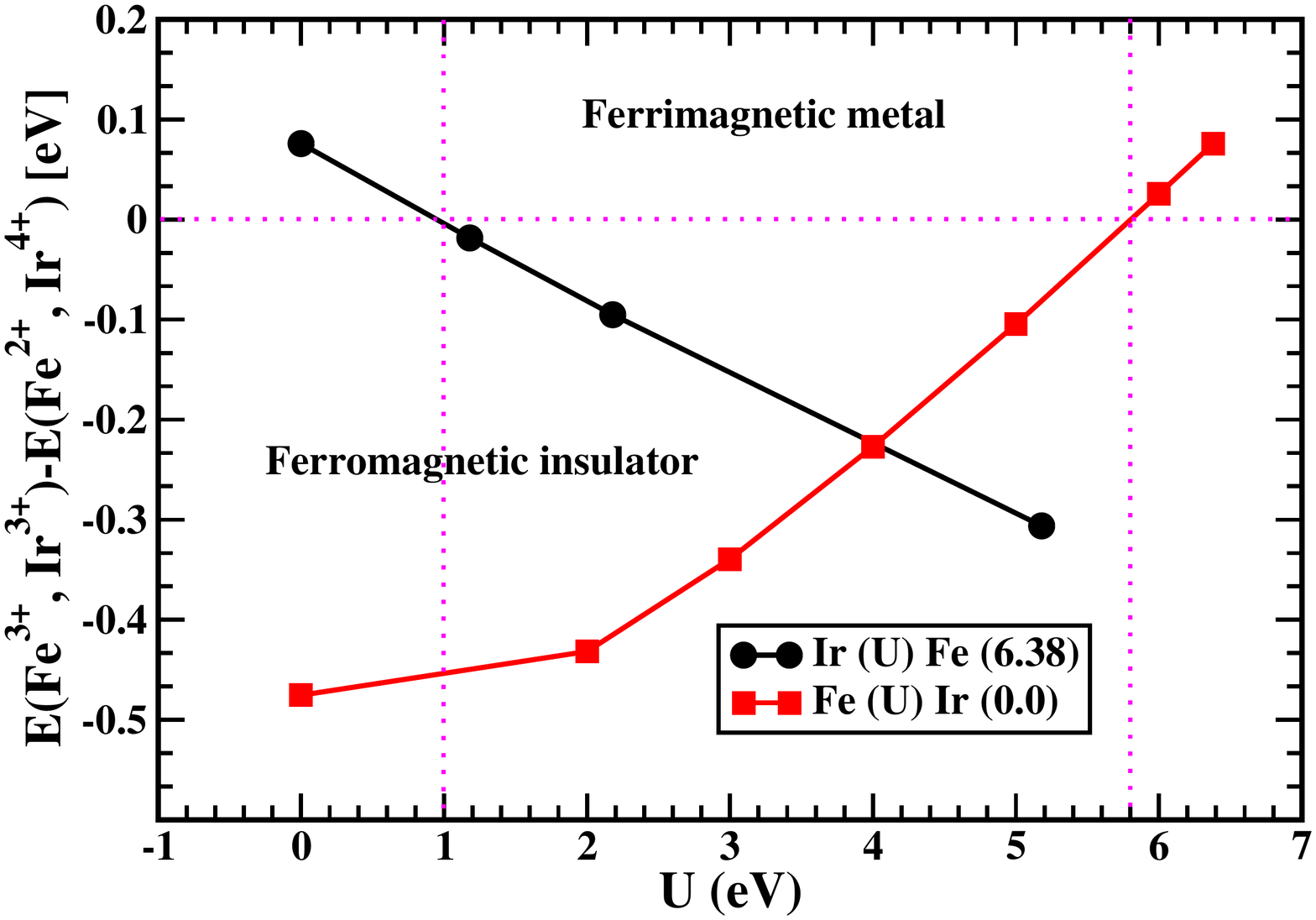}
\caption{Phase diagram of monoclinic $P2_1/n$ structure on
energy (E)-correlation ($U$) plane. There are two phases, the plane just above
the dotted horizontal line is ferrimagnet metallic and one just below it
is a ferromagnetic insulator. The solid black line represents the change in
energy by varying $U$ on Ir-sites keeping the $U$ on Fe-site fixed. Similarly,
the red line shows the change in energy by varying $U$ on Fe-sites fixing $U$ =
0 on Ir-sites.}
\label{Correlation_phase}
\end{figure}

The high $U$ value calculated for Ir is contrary to the usual expectation of lowered correlations in the more diffuse 5$d$ levels. Hence, we tested the
robustness of the ($3+$,$3+$) state by scanning the $U$ (Ir) keeping the $U$ (Fe) fixed
at 6.38 eV. The results are summarized in Figure~\ref{Correlation_phase}.
%%%%%%%%%%%%%%%%%%%%%%%%%%%%%%%%%%%%%%%%%%%%%%%%%%%%%%%%%%%%%%%%%%%%%%%%%%%%%%%%
%%%%%%%%%%%%%%%%%%%%%%%%%%%%%%%%%%%%%%%%%%%%%%%%%%%%%%%%%%%%%%%%%%%%%%%%%%%%%
We observed that, even at a lower $U$ value of 1.0 eV, the system already undergoes a transition
from ferrimagnetic (Fe$^{2+}$, Ir$^{4+}$) to ferromagnetic (Fe$^{3+}$, Ir$^{3+}$)
state. The energy of ferromagnetic states lowers further as we increase $U$
(Ir) towards our calculated value ($U$ = 5.18 eV). This result indicates that,
correlation effects at Ir play an important role in BFIO making the FiM to FM
transition a correlation driven one. We have also scanned the $U$ values for Fe-sites 
by fixing the $U$ values of Ir-sites to zero. interestingly the same phase transition
was observed at $U$ = 5.8 eV, below which the system prefers to stay in the FM
state while above it, the FiM state gets stabilized.

%%%%%%%%%%%%%%%%%%%%%%%%%%%%%%%%%%%%%%%%%%%%%%%%%%%%%%%%%%%%%%%%%%%%%%%%%%%%% 
%%%%%%%%%%%%%%%%%%%%%%%%%%%%%%%%%%%%%%%%%%%%%%%%%%%%%%%%%%%%%%%%%%%%%%%%%%%%% 
\subsection{Electronic and magnetic properties of \ce{Bi2FeIrO6}}
\begin{figure}[pbht!]
\includegraphics[trim=0.0cm 1.0cm 0.0cm
2.5cm,clip=true,width=0.480\textwidth]{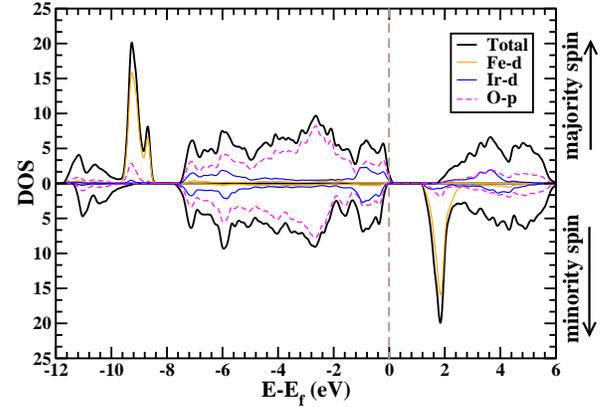}
\caption{The density of states (DOS) of FM \ce{Bi2FeIrO6}, calculated
using the GGA+$U$ method with $U_{eff}$=6.38, 5.18 eV for Fe and Ir,
respectively. The dashed vertical zero line is set to the Fermi energy. The
local magnetic moments on Ir site are antialigned with equal magnitude leading
to zero magnetic moment, while on Fe site the up spin channel is completely
filled and the down spin channel is completely empty, hence results in a net
magnetic moment in the system. The solid black line, total DOS; orange solid
line, Fe-$d$ states; blue line, Ir-$d$ states and solid magenta line, O-$p$
states.}
\label{Bi2FeIrO6_FM_Pdos}
\end{figure}

\begin{figure}[pbht!]
\includegraphics[trim=0.0cm 1.0cm 0.0cm
2.5cm,clip=true,width=0.480\textwidth]{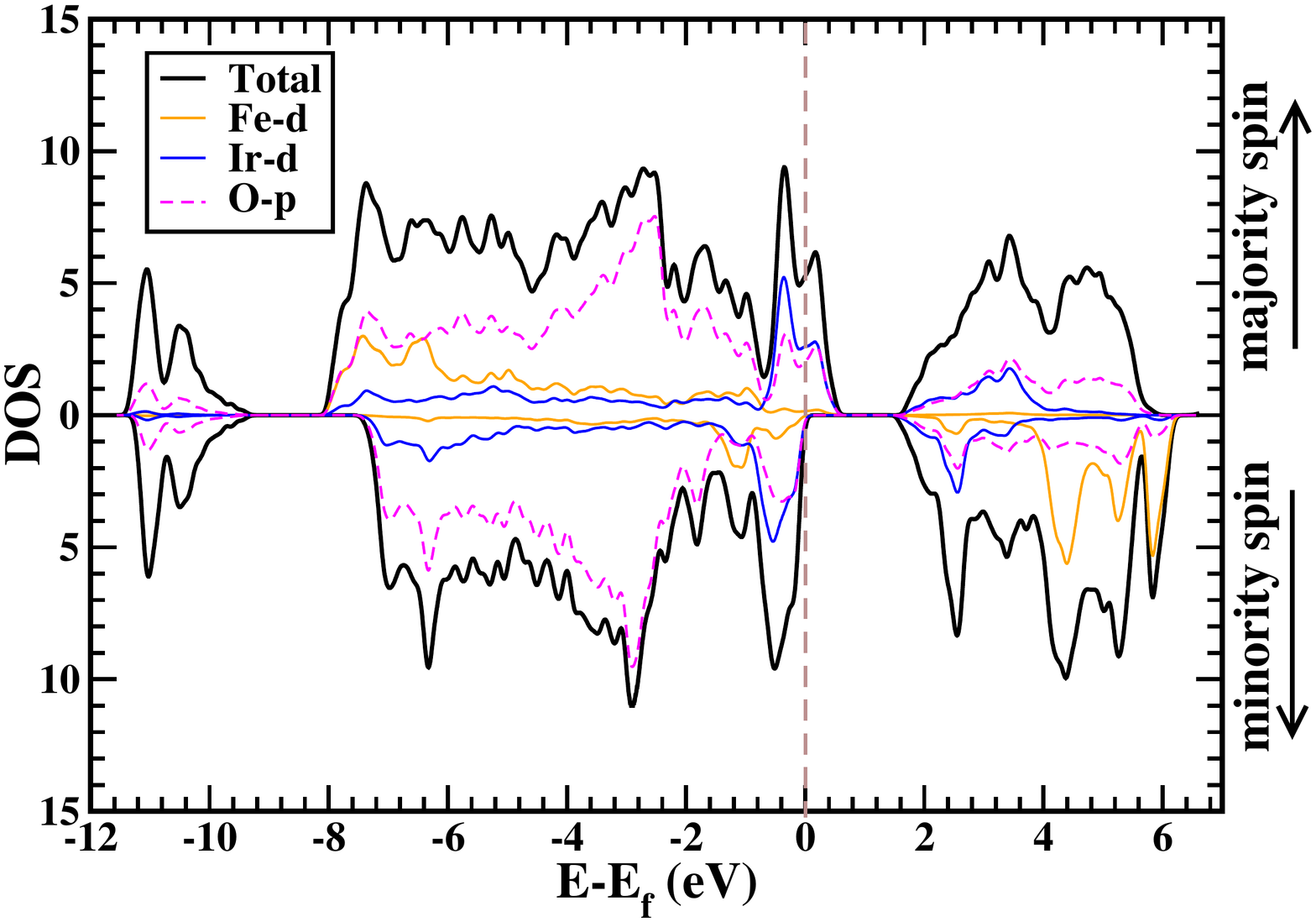}
\caption{The density of states (DOS) of G-type AFM \ce{Bi2FeIrO6},
calculated using GGA+$U$ method with $U_{eff}$=6.38, 0.0 eV for
Fe and Ir, respectively. The dashed vertical zero line is set to the Fermi
energy. The local magnetic moments on Fe and Ir are antialigned with unequal
magnitude leading to a net magnetic moment on the system. The solid black
line, total DOS; green solid line, Fe-$d$ states; blue line, Ir-$d$ states and
solid magenta line, O-$p$ states.}
\label{Ferri_PDOS}
\end{figure}

In this section, we analyze the magnetic and electronic properties of
DP \ce{Bi2FeIrO6} thin-films in detail. The DOS of both the possible 
(FM and FiM) phases are studied and are shown 
in Figure~\ref{Bi2FeIrO6_FM_Pdos} and Figure~\ref{Ferri_PDOS}, respectively. 
As is evidenced from Figure~\ref{Bi2FeIrO6_FM_Pdos}, the states close to Fermi
level (E$_F$) are dominated by Ir $d$-states strongly hybridized with O
$p$-states. The dominating part of O $p$-states separated from Fe and Ir
$d$-states are located below Fermi level within the range from -7.8 eV to -2.4
eV.
%%%%%%%%%%%%%%%%%%%%%%%%%%%%%%%%%%%%%%%%%%%%%%%%%%%%%%%%%%%%%%%%%%%%%%%%%%%%%%%%
%%%%%%%%%%%%%%%%%%%%%%%%%%%%%%%%%%%%%%%%%%%%%%%%%%%%%%%%%%%%%%%%%%%%%%%%%%%%%
In the majority spin channel, the Fe $t_{2g}$, as well as $e_g$ states, remain
filled while in Ir-sites the $t_{2g}$ states are completely filled and the $e_g$
states remain empty. In the minority spin channel, the Fe $t_{2g}$ and $e_g$
states remain completely empty while in Ir-sites the $t_{2g}$ states remain
filled and the $e_g$ states stay completely empty, thus give rise to an
insulating character in both the spin channels. This rare FM insulating nature of
BFIO makes it very special.

\begin{figure}[pbht!]
\includegraphics[trim=0.0cm 0.0cm 0.0cm
1.0cm,clip=true,width=0.48\textwidth]{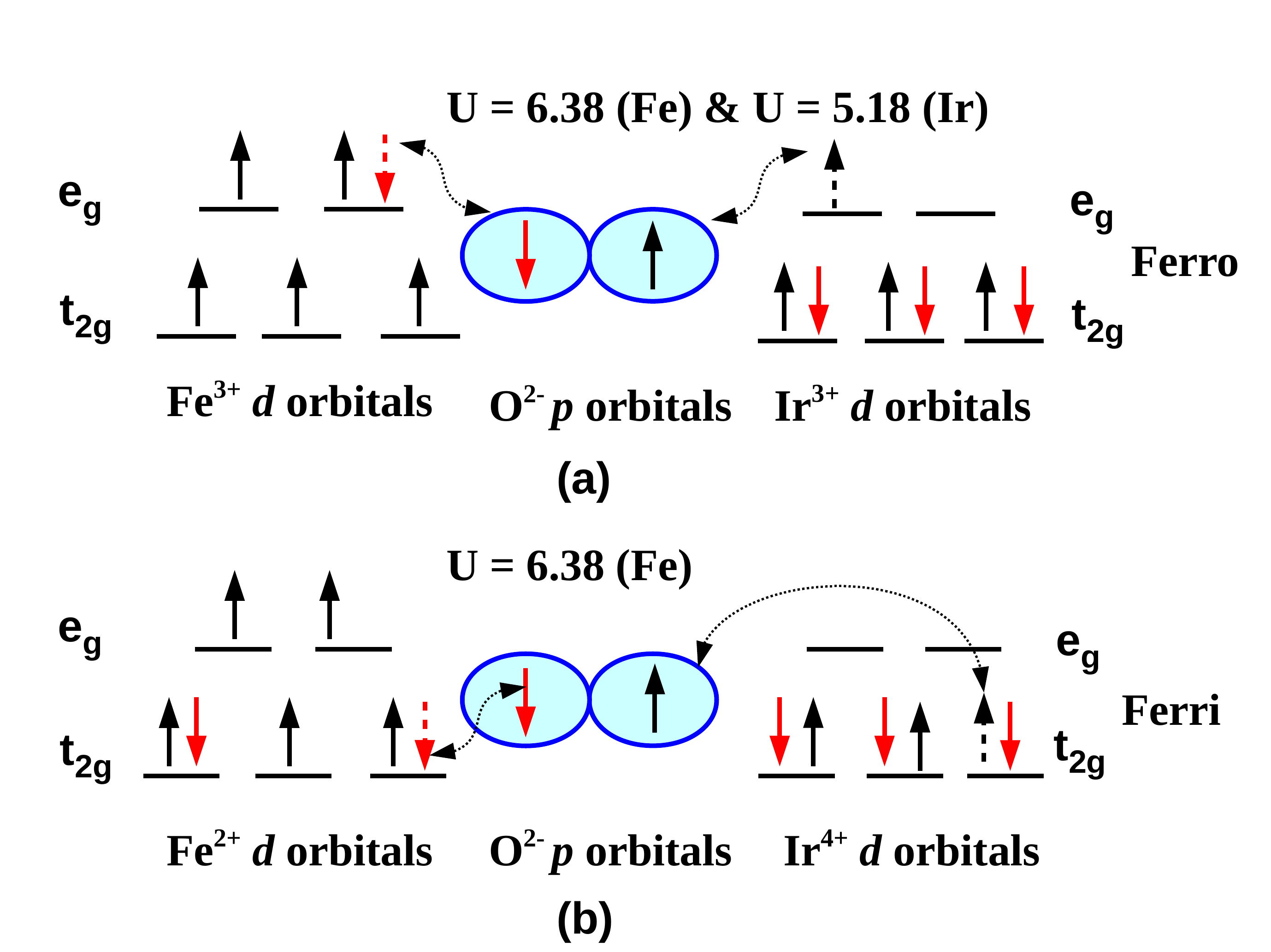}
\caption{Schematic level diagrams of 3$d$-5$d$ orbitals. (a) Schematic
representation of the 3$d$-5$d$ electrons of $d$ orbitals in Fe$^{3+}$,
Ir$^{3+}$ oxidation state. The interaction between the two transition metals
via O $p$ orbitals leads to ferromagnetism. (b) Representation of the
3$d$-5$d$ electrons of $d$ orbitals in Fe$^{2+}$, Ir$^{4+}$ oxidation state.
For case (b) the interaction between the atom leads to an AFM. The
antiferromagnetic and ferromagnetic interaction basically depends on the
occupations of the $e_g$ orbitals.}
\label{Configuration}
\end{figure}

Figure.~\ref{Ferri_PDOS} depicts the DOS of FiM metallic phase when
the on-site Coulomb repulsion is switched off on 5$d$ (Ir)-sites.
The metallic character of this state comes from the Ir majority spin
channel due to the presence of a hole in $t_{2g}$ states (see also
Figure~\ref{Configuration}). However, in the minority spin
channel, all the $t_{2g}$ states of Ir are occupied. It is noticed that the
$t_{2g}$ states of Fe and Ir co-exist around the Fermi level with a finite
hybridization between them. In this state, Ir ion is in formal oxidation
state 4+ and Fe takes high spin 2+ state, at U = 6.38 (Fe), 0 (Ir) (see
Figure~\ref{Correlation_phase}). 
%%%%%%%%%%%%%%%%%%%%%%%%%%%%%%%%%%%%%%%%%%%%%%%%%%%%%%%%%%%%%%%%%%%%%%%%%%%%%
%%%%%%%%%%%%%%%%%%%%%%%%%%%%%%%%%%%%%%%%%%%%%%%%%%%%%%%%%%%%%%%%%%%%%%%%%%%%%%
Due to these electronic configurations of outer valence shells and nearly
149$^\circ$ Fe-O-Ir angle, the magnetic moments of Fe and Ir interacts through an
antiferromagnetic superexchange mechanism satisfying the Goodenough-Kanamori
(GK) rules~\cite{good,kanamori1,Wollan}. Therefore, the system ends up with a
FiM configuration with a total magnetic moment of 1 $\mu_B$/f.u. Like
the FM insulating state, the metallic AFM states are also very special due to
their rare existence. 
%%%%%%%%%%%%%%%%%%%%%%%%%%%%%%%%%%%%%%%%%%%%%%%%%%%%%%%%%%%%%%%%%%%%%%%%%%%%%%
%%%%%%%%%%%%%%%%%%%%%%%%%%%%%%%%%%%%%%%%%%%%%%%%%%%%%%%%%%%%%%%%%%%%%%%%%%%%%%%%

As we know, 5$d$ transition metals have an intrinsically strong spin-orbit
coupling (SOC). Particularly, iridates are known to display SOC
effect~\cite{Fang,Zeb,Nie,Han,Haskel,Hua} in perovskites and DPs. Motivated by this, 
we extended our calculation to consider the SOC effect on top of GGA 
as well as GGA+$U$ functionals. We observed that
adding SOC to GGA, lowered the band gap, while the insulating behavior was
recovered by adding $U$. However, there is no major change in the magnetic
properties of the system. Once again the system preferred to stay in
the Fe$^{3+}$, Ir$^{3+}$ oxidation state even with SOC. The low spin 3+ oxidation 
state of Ir ($d^6$) ion, weakens the role of SOC in BFIO. 

As can be seen in Fig.~\ref{Bi2FeIrO6_FM_Pdos}, the conduction band is made up almost entirely by Fe states whereas valence
band results from Ir-O hybridised states. In principle, this implies that the photoexcitation can lead to electron-hole separation in the sublattices which is a desirable feature for photovoltaic applications.

\section{Summary}
In summary, we have designed two Bi-based double-perovskites \ce{Bi2FeReO6}
and \ce{Bi2FeIrO6} mimicking the thin-film geometry grown along [001] direction.
These two structures were constructed by keeping the Fe atom
fixed at B-site and substituting Re and Ir (5$d$) atoms in B$^\prime$-sites of
ordered DP A$_{2}$BB$^\prime$O$_{6}$ structure. The calculated tolerance factors
indicate that both the compounds will adopt perovskite
(orthorhombic/monoclinic) crystal structure. The \ce{Bi2FeReO6} thin-film
becomes stable adopting a monoclinic $P2_1/n$ phase under a wide range of
ES. By incorporating various types of magnetic ordering we show that
the thin-film \ce{Bi2FeReO6} prefers a C-AFM spin ordering in its
ground-state with zero magnetic moments. 
However, at -2.7\% strain the structure adopts G-AFM ordering through
a monoclinic to monoclinic structural transition. The G-AFM phase of BFRO
provides a magnetic moment of 1 $\mu_B$/f.u. which is comparable to the
previously reported  Bi-based 3$d$-5$d$ compounds~\cite{Marjana}. Our DOS calculations
indicate that the \ce{Bi2FeReO6} is an insulator with a band gap of
around 1.2 eV in thin film form. We find the correlation effect on Re (5$d$)-site to be
very important, as on-site Coulomb repulsion ($U$) widens the band gap significantly. 
In this compound, Fe takes a high spin 2+ oxidation state, whereas Re adopts its formal high spin 4+ state. 
One important finding of our calculations is that the oxygen octahedra around Fe atoms undergo a 
distortion leading to long-short Fe-O bonds in the octahedra, which is
not observed in the BFIO thin-film. 

In \ce{Bi2FeIrO6} the $P2_1/n$ symmetry structure emerged as the
ground state. Both the transition metals (Fe, Ir) takes 3+ oxidation state,
where Fe acquires high spin configuration and Ir prefers to stay in a low spin.
This configuration of \ce{Bi2FeIrO6}  leads to a huge
ferromagnetic ground state with a magnetic moment of 5 $\mu_B$/f.u. throughout
the considered ES. However, when the onsite Coulomb repulsion term
on Ir-site is switched off, the system stabilizes in FiM ground state
with a magnetic moment of 3 $\mu_B$/f.u. over the ES. The
DOS calculations show that the FM state is an insulator
while the FiM state is a half-metal. Our calculations show that even a
small value of $U$ on Ir site could, in principle, switch the (${2+}$,
${4+}$) oxidation state to (${3+}$, ${3+}$) state. So, there is a
correlation driven metal to insulator transition in BFIO
thin-film under ES. Given that ferromagnetic insulators and antiferromagnetic metals 
are very rare, the \textit{ab initio} designed BFIO could prove to be quite technologically
significant.

Both the compounds can be used in spintronic applications due to their magnetic
and half-metallic nature. Apart from this, their narrow band-gaps
($\sim$ 1.2 and 1.3, respectively) make them a suitable candidate for PV applications. Although
these materials are not ferroelectric, an appropriate
doping strategy can make them ferroelectric~\cite{rout2018epitaxial,zhao2014}. Apart
from PV applications, these materials can also be used for the photocatalytic
activity, such as solar water splitting and water purification. Like the parent BFO
compound, the narrow band gap nature of these 3$d$-5$d$ compounds could allow
them to harvest parts of the visible light of the solar spectrum. We hope that our observations will initiate further
experimental efforts to verify the interesting predictions made in this work.

\acknowledgements{The authors would like to thank Dr. D. S. Rana at IISER
Bhopal for helpful discussions. The authors
gratefully acknowledge IISER Bhopal for computational resources and funding. PCR
is thankful to Council of Scientific and Industrial Research, India, for the
research fellowship.}

\bibliographystyle{apsrev4-1}
\bibliography{reference}
\end{document}